\documentclass{article}
\usepackage{graphicx} 
\usepackage{amsfonts} 
\usepackage{amsmath} 
\usepackage{placeins}
\usepackage{amsmath,amssymb}
\usepackage{booktabs}
\usepackage{placeins}
\usepackage{graphicx}
\usepackage{hyperref}
\usepackage{subfig}
\usepackage{xcolor}
\usepackage{caption}
\usepackage[a4paper,
    left=2cm,
    right=2cm
]{geometry}

\title{Analysis of a Core Envelope Model for PSR J1614-2230 in $f(R)= R+\alpha R^{2} +\gamma R^{3}$ gravity and Tidal Deformability}
\author{
A. Khalid \thanks{ayeshakhalid4238@gmail.com; kha0057@vsb.cz}${}^{~(a,b)}$,
Muhammad Bilal Riaz  \thanks{muhammad.bilal.riaz@vsb.cz; bilalsehole@gmail.com }${}^{~(a,c)}$,
S. A. Mardan \thanks{syedalimardanazmi@yahoo.com}${}^{~(c,e)}$,
M. Inç\thanks{minc747800@gmail.com}${}^{~(d)}$,\\
${}^{a}$IT4Innovations, VSB – Technical University of Ostrava, \\Ostrava, Czech Republic.\\
${}^{b}$ Faculty of Electrical Engineering and Computer Science, \\VSB – Technical University of Ostrava, Ostrava, Czech Republic\\
${}^{c}$ Department of Mathematics, Firat University, Elazig,\\ 23119, Turkiye\\
${}^{d}$Department of Mathematics, Khazar University, Baku, Azerbaijan.\\
${}^{e}$ Department of Mathematics, University of Management and Technology,\\  Lahore, Pakistan.\\
}
\date{October 2026}
\begin{document}
\maketitle

\begin{abstract}
In this paper, we study the tidal deformability of anisotropic core envelope model for the compact star PSR J1614$-$2230 in the framework of modified $f(R)=R+\alpha R^2+\gamma R^3$ gravity. The stellar interior is divided into two regions, where the core is model by a linear equation of state (EoS), and the envelope follows a quadratic EoS. The Tolman VII form is used for radial metric potential in both interior regions. The solution is matched at the core envelope boundary as well as with the exterior Schwarzschild spacetime using the junction conditions. The tidal Love number is obtained by solving the coupled perturbative equations for the scalar field. The developed core envelop model is free from singularities, and also satisfies energy conditions, causality conditions, adiabatic index constraint, Herrera cracking condition and hydrostatic equilibrium balancing force criteria. We conclude that our core–envelope model is physically viable in cubic $f(R)$ gravity and has a tidal deformability consistent with GW170817.  
\end{abstract}

\section{Introduction}
\label{sec:1}

Neutron stars are very dense astronomical objects that exist in the universe as stellar residues of supernovae massive stars, when they undergoes a gravitational collapse by burning out all of its interior nuclear fluid. The neutron stars are highly dense that a teaspoon of its mass can be compared to mass of earth. In this regard, mathematical models of these compact objects, which are consistent with observational data plays a key role in understanding their internal structure. PSR J1614$-$2230 is one of the known heaviest neutron stars and has become an important observational testing standard to study the properties of matter in extreme gravitational environments. Demorest et al.~\cite{PSR1} measured the mass of PSR J1614$-$2230 to be $(1.97\pm0.04)\,M_\odot$ by using Shapiro delay. Fonseca et al.~\cite{PSR2} obtained a more precise mass value by observing  pulsar's nine years of data as $(1.928\pm0.017)\,M_\odot$. Nazar et al.~\cite{PSR3} modeled PSR J1614$-$2230 as an anisotropic compact star in a modified gravity theory. More recently, Mendes et al.~\cite{PSR4} provided mass-radius constraints on PSR J1614$-$2230 subject to EoS for dense matter. Mauviard et al.~\cite{PSR5} measured its radius to be about $10$~km, confirming that PSR J1614$-$2230 is both massive and highly compact. The internal structure of a star depends not only on the density of its matter but also on the pressure distribution. The difference in radial and tangential pressures gives rise to anisotropic behavior, and strongly affects its equilibrium state.\\

In massive stellar objects, pressure anisotropy adds another degree of freedom in the analysis of the internal equilibrium. Inclusion of  anisotropy provides a more general context for observing stellar configurations in the framework of the anisotropic fluid. Recently,  Das et al. \cite{AN1,AN3} studied anisotropic compact star models under the vanishing complexity condition and with quasi-local anisotropy conditions. Maurya et al. \cite{AN2} obtained an analytical solution for self-bound stellar anisotropic systems in general relativity. Khunt et al. \cite{AN4} studied that the influence of anisotropy on the internal configuration of neutron stars and conclude that it can enhance mass upto 20\%. Mathias and Mkenyeleye \cite{AN5} proposed a simple relativistic model of an anisotropic star, which provides a convenient starting point for studying how anisotropy shapes stellar structure. These studies motivates us to in-cooperate anisotropic effects in the study of but stars. However, along with anisotropy suitable gravitational framework choice also plays a key role in the interior analysis of stars. In extreme gravitational environments modified gravitational theories are more suitable for understand of stars.\\

General relativity has gained considerable success in the explanation of weak field regimes but the problems associated with cosmic acceleration, cosmological constant, quantum corrections and strong gravitational field regime motives the mitigation towards modified gravity. The $f(R)$ framework introduced by Sotiriou \cite{Sotiriou2010}, extends the Einstein-Hilbert action by introducing nonlinear curvature contributions, offering a way to investigate gravitational effects beyond the standard Einstein framework. Capozziello et al. \cite{FR1} studied the neutron star mass–radius relations for neutron stars in the context of \(f(R)\) gravity. Yousaf et al. \cite{FR2} discussed the influence of \(f(R)\) theory selection in self-gravitation anisotropic systems. Several recant studies highlights the relation of \(f(R)\) theory in modeling of compact objects under various constraints \cite{FR3,FR4,FR5}. These developments pave the way for studying curvature effects in the stars where the matter can change significantly from one region to another. In this regard, the core-envelope models provide a natural way to explore the interaction between different matter distributions and modified gravity in this context, while preserving the stellar structure of compact objects.\\

In understanding the internal structure of a star, the core-envelope approach offers a more detailed perspective as it treats the inner  core and outer envelope separately. This two layered system enables variations in the composition of the matter without losing the continuity of the stellar configuration. In the past few years the development of core–envelope models comes to high spotlight under different constraints due to their offering of further deeper insides of compact objects \cite{CE1,CE2,CE3,CE4,CE5}. In the development of mathematical models a suitable choice of EoS is essential to close the system of field equations. Different models of compact objects based on various EoS were presented by many researchers to highlight the importance of EoS selection in order to describe their internal structure of stars. In literature many different types of EoSs were used by researchers in models development like linear, quadratic, polytropic and many others \cite{QEoS1,QEoS2,QEoS3,QEoS4,QEoS5,LEoS1,LEoS2,LEoS3,LEoS4}. In mathematical modeling of compact objects, it is very important to test it stability under the influence of external forces exerted on the star. In this regard tidal deformation is key factor to understand external influences on the equilibrium of star. \\

Tidal deformability measures how easily a dense star is deformed by an external gravitational field, and reveals the crucial details of stellar interior that cannot be observed by equilibrium properties alone. Mohanty et al.~\cite{TD1} showed that anisotropy noticeably changes the mass, radius and tidal deformability of neutron stars. The tidal deformability in higher curvature and higher dimensional conditions were discussed in \cite{TD3,TD4,TD5}. Together, these studies show that the tidal response depends on both the stellar matter and the theory of gravity. This response is quantified by the tidal Love number, which measures the deformation induced by the tidal field. The Love number is more sensitive than mass and radius to the matter distribution inside the star. Yazadjiev et al.~\cite{TLN1} computed tidal Love numbers in $f(R)$ gravity. Yusmantoro et al.~\cite{TLN3} obtained mass-radius relations and Love numbers in modified gravity. More recently, Pani et al.~\cite{TLN5} went beyond the linear approximation and studied the nonlinear tidal response of compact stars. However, treat the star as a single layer, and the tidal response of a core-envelope star in $f(R)$ gravity has not yet been examined.\\

In this work, we provide a novel the tidal deformability approach for deeper understanding of PSR J1614-2230 using a core–envelope model in cubic $f(R) = R + \alpha R^{2} + \gamma R^{3}$ gravity. The core of the star is modeled through a linear EoS, and envelope via a quadratic EoS. This combination of a linear–quadratic EoS in core–envelope model in cubic $f(R)$ gravity with tidal deformability has not been analyzed before for understanding of compact stars. 
It will be interesting to observe this two layered interior arrangement under the influence modified $f(R)$ gravity. Tidal deformability measures how much a star is deformed by an external gravitational field, and its value depends on how matter is distributed inside the star. Tidal deformability also gives information about the stellar interior beyond what mass and radius provide. Gravitational wave events also measure tidal deformability, which allows us to test whether a layered star in cubic $f(R)$ gravity is consistent with the GW170817 bound.\\


The layout of this manuscript is as follows. In Section~\ref{sec:2} the $f(R)$ field equations are presented for an anisotropic fluid to obtain the solutions for the core and envelope. Section~\ref{sec:3} focus on the continuity of core with envelop and exterior region. 
In Section~\ref{sec:4}, we check the physical acceptability of the solution through its metric potentials, matter profiles, anisotropy, energy conditions, causality, stability and equilibrium, and then calculate the tidal Love number and tidal deformability of the star. In the last, section~\ref{sec:5} we summarizes the main results.

\section{Modified Field Equations}
\label{sec:2}
Owe start with Einstein-Hilbert action \cite{TLN1}, where the Ricci scalar $R$ is replaced by an arbitrary function $f(R)$
\begin{equation}
S=\int d^{4}x\sqrt{-g} \left[\frac{f(R)}{2\kappa}+\mathcal{L}_{m}\right],
\label{eq:1}
\end{equation}
here $\kappa = 8\pi G$ and that $G = c = 1$ is used in geometrized units. Relative to $g^{ij}$ in the metric formalism leads to the field equations
\begin{equation}
R_{ij}f_R-\frac{1}{2}f(R)g_{ij} +\left(g_{ij}\Box-\nabla_i\nabla_j\right)f_R =\kappa T_{ij},
\label{eq:2}
\end{equation}
in which $f_R\equiv df/dR$, $\nabla_i$ is the covariant derivative, $\Box\equiv g^{ij}\nabla_i\nabla_j$ is the d'Alembert operator and $T_{ij}$ is the energy momentum tensor of ordinary matter. The compact form of Eq.~(\ref{eq:2}) is given by
\begin{equation}
G_{ij}=\frac{\kappa}{f_R} \left(T_{ij}^{(D)}+T_{ij}\right) =T_{ij}^{(eff)},
\label{eq:3}
\end{equation}
where $G_{ij}=R_{ij}-\frac{1}{2}Rg_{ij}$ is the Einstein tensor and
\begin{equation}
T_{ij}^{(D)}=\frac{1}{\kappa} \left\{ \nabla_i\nabla_j f_R-\Box f_R g_{ij} +\left(f-Rf_R\right)\frac{g_{ij}}{2} \right\}.
\label{eq:4}
\end{equation}
The tensor $T^{(D)}_{ij}$ in-cooperates every contribution that comes from the extra curvature terms and together with $T_{ij}$ it makes up the effective energy-momentum tensor $T^{(eff)}_{ij}$.\\
To model the interior of a compact star, we use a static and spherically symmetric line element,
\begin{equation}
ds^{2}=e^{\nu(r)}dt^{2}-e^{\lambda(r)}dr^{2}
-r^{2}\left(d\theta^{2}
+\sin^{2}\theta\,d\phi^{2}\right),
\label{eq:5}
\end{equation}
where the metric potentials $\nu$ and $\lambda$ depend only on the radial coordinate $r$ and anisotropic energy momentum tensor is given by
\begin{equation}
T_{ij}=(p_t+\rho)u_i u_j
+(p_r-p_t)\chi_i\chi_j-p_tg_{ij},
\label{eq:6}
\end{equation}
with
\[
u^i=e^{-\nu(r)/2}\delta^i_0,
\qquad
\chi^i=e^{-\lambda(r)/2}\delta^i_1.
\]
using the Eqs. (\ref{eq:5}) and  (\ref{eq:6}) into Eq. (\ref{eq:3}), we obtain the following expressions for the
matter variables
\begin{subequations}\label{eq:7}
\begin{align}
\rho &= \frac{e^{-\lambda}}{2r^{2}}\Big(
-2r^{2}f_R''
+\lambda'r^{2}f_R'
-4rf_R'
-2\lambda'rf_R
+2f_R-2f_Re^{\lambda}
+r^{2}e^{\lambda}Rf_R
-r^{2}fe^{\lambda}
\Big),
\label{eq:7a}
\\[6pt]
p_r &= \frac{e^{-\lambda}}{2r^{2}}\Big(4rf_R'
+\nu'r^{2}f_R'
+2e^{\lambda}f_R
-2f_R-r^{2}f_Re^{\lambda}R
+r^{2}e^{\lambda}f
+2rf_Rr\nu'
\Big),
\label{eq:7b}
\\[6pt]
p_t &= \frac{e^{-\lambda}}{4r}\Big(-2rf_{R}\nu''
-\nu'\lambda'f_{R}r
-(\nu')^{2}f_{R}r
+2f_R\lambda'+4f_R'
-2\nu'f_{R}
+2\nu'f_R'r
-2\lambda'rf_R'
\nonumber\\
&+4f_R''r
+2e^{\lambda}fr
-2e^{\lambda}f_{R}rR
\Big),
\label{eq:7c}
\end{align}
\end{subequations}
where prime represents differentiation with respect to $r$. The anisotropy inside the star is measured by $\Delta = p_t-p_r$. 
Regularity at the center requires $\Delta$ to vanish at $r=0$. Note that Eqs.~(\ref{eq:7a})-(\ref{eq:7c}) contain more unknowns than number of equations, so we need a way out to close the system. In order to do so, we introduce the transformations in next subsection~\ref{2.1}.

\subsection{Transformations for Modified Field Equations}\label{2.1}
We use change of variable technique here to handle system of Eqs.~(\ref{eq:7a})-(\ref{eq:7c}) by introducing following transformation
\[
x=r^{2}, \qquad
y(x)=e^{\nu(r)}, \qquad
z(x)=e^{-\lambda(r)},
\]
by using above transformation and chain rule the derivative changed as $dx/dr=2r=2\sqrt{x}$,~~
$d/dr=2\sqrt{x}\,d/dx$. From this point on, a prime represents differentiation along $x$. Under the transformation, Eqs.~(\ref{eq:7a})-(\ref{eq:7c}) become
\begin{subequations}\label{eq:8}
\begin{align}
\rho &= \Bigg[-f_R''Z
-\frac{1}{x}
+\frac{2(Z-Z')f_R'}{\sqrt{x}}
-\frac{(Z'-1+R)f_R}{2}
\Bigg],
\label{eq:8a}
\\[6pt]
p_r &= \Bigg[\left(
\frac{Zf_R}{\sqrt{x}}
+\frac{Zf_R'}{2}
\right)\frac{y'}{y}
+\frac{2Zf_R'}{\sqrt{x}}
+\left(\frac{1-Z}{x}
-\frac{R-1}{2}
\right)f_R
\Bigg],
\label{eq:8b}
\\[6pt]
p_t &= \Bigg[-\frac{Zf_Ry''}{2y}
+\frac{Zf_R}{4}
\left(\frac{y'}{y}\right)^2
+\left(
\frac{Zf_R'}{2}
-\frac{Z'f_R}{4}
-\frac{Zf_R}{2\sqrt{x}}
\right)\frac{y'}{y}
+\left(Z+\frac{Z'}{2}\right)f_R''+\left(
\frac{1-R}{2}
+\frac{Zf_R'}{\sqrt{x}}
-\frac{Z'}{2\sqrt{x}}
\right)f_R
\Bigg].
\label{eq:8c}
\end{align}
\end{subequations}
In this form, the field equations are easier to use for computing the interior properties of the star.

\subsection{Core region solution}

To describe the inner core of the star, we need to specify the metric functions and simplify the system with a linear EoS. For the metric functions, we take a simple Tolman VII \cite{TVII} polynomial in $x$ as
\begin{equation}
Z=e^{-\lambda_C}=bx^2-ax+1,\qquad p_{r_C}=\xi\rho+\beta,
\label{eq:9}
\end{equation}
where $a$, $\xi$, $b$, and $\beta$ are constants and the subscript $C$ refers to the core. Since $Z(0)=1$, the metric function is well behaved at the center. Also, the function of $f(R)$ in this work is defined as
\begin{equation}
f(R)= R+\alpha R^{2} +\gamma R^{3},
\label{eq:10}
\end{equation}
with $\gamma$ and $\alpha$ constant. The quadratic term has the familiar Starobinsky form, and the cubic term adds a further correction that becomes important in regions of high curvature, such as the interior of a dense star. General relativity is recovered for $\alpha=\gamma=0$. For this model the quantities appearing in the field equations are 
$f_R=1+2\alpha R+3\gamma R^2$~~, $f'_R=(2\alpha+6\gamma R)R'$ and
$f''_R=(2\alpha+6\gamma R)R''+6\gamma(R')^2$. Inserting the EoS into Eq.~(\ref{eq:8b}) and solving for $y'/y$ for the metric function. A single integration then gives $y$ itself
\begin{equation}
\frac{y'}{y}=\left[\frac{2\sqrt{x}\left(\xi\rho_c+\beta-\frac{2(1+bx^2-ax)f_R'}{\sqrt{x}}-\left(\frac{1}{2}+\frac{1}{x}-\frac{1+bx^2-ax}{x}
-\frac{R}{2}\right)f_R\right)}{(2f_R+\sqrt{x}f_R')
(1-ax+bx^2)}\right],
\label{eq:11}
\end{equation}
In the second expression, we have used $(1-Z)/x=a-bx$ to simplify the coefficient of $f_R$. Also with Eq.~(\ref{eq:10}), the physical quantities in the core follow directly from Eqs.~(\ref{eq:8a})-(\ref{eq:8c}) takes the form
\begin{subequations}\label{eq:12}
\begin{align}
\rho_{C}&=-(1-ax+bx^2)
\left[(2\alpha+6\gamma R)R''
+6\gamma(R')^2\right]
+\frac{2(1+a-ax+bx^2-2bx)
(2\alpha+6\gamma R)R'}{\sqrt{x}}
\nonumber\\
&-\frac{(2bx-a+R-1)
(1+2\alpha R+3\gamma R^2)}{2}
-\frac{1}{x},
\label{eq:12a}
\\[8pt]
p_{r_{C}}&=\xi\rho_{C}+\beta,
\label{eq:12b}
\\[8pt]
p_{t_{C}}&=
-\frac{(1+bx^{2}-ax)
(1+2\alpha R+3\gamma R^2)y''}{2y}+\frac{(1+bx^{2}-ax)
(1+2\alpha R+3\gamma R^2)}{4}
\left(\frac{y'}{y}\right)^2
\nonumber\\
&+\Bigg[
\frac{(1-ax+bx^2)
(2\alpha+6\gamma R)R'}{2}-\frac{(2bx-a)
(1+2\alpha R+3\gamma R^2)}{4}-\frac{(1-ax+bx^2)
(1+2\alpha R+3\gamma R^2)}{2\sqrt{x}}
\Bigg]\frac{y'}{y}
\nonumber\\
&+\Bigg[
\frac{1-R}{2}
+\frac{(1-ax+bx^2)
(2\alpha+6\gamma R)R'}{\sqrt{x}}-\frac{2bx-a}{2\sqrt{x}}
\Bigg](1+2\alpha R+3\gamma R^2)
\nonumber\\
&
+\left(1-ax+bx^2+\frac{2bx-a}{2}\right)
\Big[(2\alpha+6\gamma R)R''+6\gamma(R')^2\Big].
\label{eq:12c}
\end{align}
\end{subequations}
The density $\rho$ appearing in Eq.~(\ref{eq:11}) is the core density $\rho_C$ given by Eq.~(\ref{eq:12a}). The tangential pressure needs both $y'/y$ and $y''/y$, and both can be read off from Eq.~(\ref{eq:11}) and its derivative. The anisotropy for core is formulated by $\Delta _{C} = p_{t_{C}}-p_{r_{C}}$. Lastly, we need the Ricci scalar in the core. Written in terms of the radial
coordinate, it is
\begin{equation}
R=\left[-2br^{2}-2b+\frac{4(a-br^{2})}{r}+2a-\frac{(a-2br^{2})^{2}}{1-ar^{2}+br^{4}}\right].
\label{eq:13}
\end{equation}
with Ricci scalar defined above the equations for core are complete and in next subsection, we will provide equations for envelop.

\subsection{Envelope region solution}\label{2.3}

We now discuss the envelope, the outer layer that lies between the core and the stellar surface. The matter here is less dense than in the core, and a linear EoS is unlikely to describe it well over the whole layer. We keep the same Tolman VII \cite{TVII} form for the metric potential and quadratic EoS.
\begin{equation}
Z=e^{-\lambda_E}=bx^2-ax+1,\qquad p_{r_E}=A \rho^{2}+B \rho -O,
\label{eq:14}
\end{equation}
where $A$, $B$ and $O$ are constants and the subscript $E$ denotes the envelope. Depending on the sign and size of $A$, it can make the matter stiffer or softer as the density changes. Substituting the EoS~(\ref{eq:14}) into Eq.~(\ref{eq:8b}) and proceeding
as in the core, we obtain the equation for the metric function $y$ and its formal solution
\begin{equation}
\frac{y'}{y}
=\frac{1}{\mathcal{J}(x)}
\Bigg[
A\rho_E^2+B\rho_E-O
-\frac{2(1-ax+bx^2)(2\alpha+6\gamma R)R'}
{\sqrt{x}}-\left(a-bx+\frac{1-R}{2}\right)
(1+2\alpha R+3\gamma R^2)
\Bigg],
\label{eq:15}
\end{equation}
where $y(x)=\exp\left[ \int\frac{y'}{y}\,dx \right]$ and the function in the denominator of Eq. (\ref{eq:15}) is \[ \mathcal{J}(x)= (1-ax+bx^2) \left[ \frac{1+2\alpha R+3\gamma R^2}{\sqrt{x}} +\frac{(2\alpha+6\gamma R)R'}{2} \right]. \] Writing the denominator as $\mathcal{J}(x)$ keeps the expression compact and makes it clear that it is simply $Z(f_R/\sqrt{x}+f'_R/2)$ evaluated for the present model. As in the core, the integral cannot generally be done in closed form and has to be evaluated numerically. The matter variables in the envelope are
\begin{subequations}\label{eq:16}
\begin{align}
\rho_{E}&=-(1-ax+bx^2)
\left[(2\alpha+6\gamma R)R''
+6\gamma(R')^2\right]
+\frac{2(1+a-ax+bx^2-2bx)
(2\alpha+6\gamma R)R'}{\sqrt{x}}
\nonumber\\
&-\frac{(2bx-a+R-1)
(1+2\alpha R+3\gamma R^2)}{2}
-\frac{1}{x},
\label{eq:16a}
\\[8pt]
p_{r_{E}}&=A \rho_{E}^{2}+B \rho_{E} -O
\label{eq:16b}
\\[8pt]
p_{t_{E}}&=-\frac{(1+bx^2-ax)
(1+2\alpha R+3\gamma R^2)y''}{2y}+\frac{(1+bx^2-ax)
(1+2\alpha R+3\gamma R^2)}{4}
\left(\frac{y'}{y}\right)^2
\nonumber\\
&+\Bigg[
\frac{(1+bx^2-ax)
(2\alpha+6\gamma R)R'}{2}
-\frac{(2bx-a)
(1+2\alpha R+3\gamma R^2)}{4}
-\frac{(1+bx^2-ax)
(1+2\alpha R+3\gamma R^2)}{2\sqrt{x}}
\Bigg]\frac{y'}{y}
\nonumber\\
&+\Bigg[
\frac{1-R}{2}
+\frac{(1-ax+bx^2)
(2\alpha+6\gamma R)R'}{\sqrt{x}}
-\frac{2bx-a}{2\sqrt{x}}
\Bigg](1+2\alpha R+3\gamma R^2)
\nonumber\\
&+\left(1-ax+bx^2+\frac{2bx-a}{2}\right)
\Big[(2\alpha+6\gamma R)R''
+6\gamma(R')^2\Big].
\label{eq:16c}
\end{align}
\end{subequations}
Together with the core solution, these expressions give a complete solution of the stellar interior. The anisotropy in envelope region is defined as $\Delta_{E}{}=p_{t_{E}}-p_{r_{E}}$. The constants of the two regions can then be fixed by requiring a smooth junction at the core-envelope interface and by matching the envelope to the exterior spacetime at the boundary, where the radial pressure must vanish.

\section{Core, envelope and exterior metric}
\label{sec:3}
The stellar model, we consider has two interior layers, a core and an envelope, surrounded by empty space. Each region has its own line element. The core occupies $0\le r\le R_C$ and the envelope $R_C\le r\le R_E$, where $R_E$ is the radius of the star. Outside the star ($r\ge R_E$) we take the Schwarzschild geometry. The three metrics are
\begin{subequations}\label{eq:17}
\begin{align}
ds^2\big|_C&
=e^{\nu_C(r)}dt^2-e^{\lambda_C(r)}dr^2
-r^2(d\theta^2+\sin^2\theta\,d\phi^2),
\label{eq:17a}
\\[6pt]
ds^2\big|_E
&=e^{\nu_E(r)}dt^2-e^{\lambda_E(r)}dr^2
-r^2(d\theta^2+\sin^2\theta\,d\phi^2),
\label{eq:17b}
\\[6pt]
ds^2\big|_B
&=-\left(1-\frac{2M}{R}\right)^{-1}dr^2
+\left(1-\frac{2M}{R}\right)dt^2-r^2(d\theta^2+\sin^2\theta\,d\phi^2),
\label{eq:17c}
\end{align}
\end{subequations}
where $M$ is the total mass of the star. For the model Eq.~(\ref{eq:10}), we have $f(0)=0$ and $f_R(0)=1$, so any vacuum spacetime with $R=0$, Schwarzschild included, solves the field equations. The exterior solution is therefore the same as in general relativity. The three regions must join smoothly. At the core, envelope and surface interface $r=R_C$, the metric functions and the radial pressure must be continuous. 
\begin{equation}
e^{\nu_C}(R_C) = e^{\nu_E}(R_C), \qquad
e^{\lambda_C}(R_C) = e^{\lambda_E}(R_C),
\qquad
p_{r_C}(R_C) = p_{r_E}(R_C).
\label{eq:18}
\end{equation}
Since $\lambda_C$ and $\lambda_E$ are given by the same function, condition Eq.~(\ref{eq:18}) is satisfied automatically. Condition Eq.~(\ref{eq:18}) fixes the integration constant that appears when $y$ is obtained in each region. At the surface $r=R_E$, the envelope has to join the exterior Schwarzschild's metric:
\begin{equation}
e^{\nu_E}(R_E) =
\left(1-\frac{2M}{R_E}\right),
\qquad
e^{\lambda_E}(R_E) =
\left(1-\frac{2M}{R_E}\right)^{-1},
\qquad
p_{r_E}(R_E) = 0.
\label{eq:19}
\end{equation}
In general relativity these conditions are enough. The $f(R)$ field equations contain derivatives of the curvature, and the junction conditions become stricter. Besides continuity of the metric, they require continuity of the extrinsic curvature $K_{ij}$, of its trace $K$ and of the Ricci scalar across the matching hypersurface \cite{TLN8}. 
\begin{equation}
 [g_{ij}]_-^+ = 0,\qquad f_R[K_{ij}]_-^+ = 0,\qquad [K]_-^+ = 0,\qquad [R]_-^+ = 0. \label{eq:20}
\end{equation}
Applying these conditions determines several of the model constants in terms of the mass, the radii and remaining free parameters. Parameters involved in equations are calculated as
\[
\begin{aligned}
a &= \frac{2M}{R_E^3}+bR_E^2,\qquad
\beta=(S_3-S_4)
\left[A(S_3-S_4)+B-\xi\right]-(S_1-S_2)
\left[A(S_1-S_2)+B\right],\\[8pt]
S_1 &=
-(1-aR_E^2+bR_E^4)
\left[(2\alpha+6\gamma R)
 R''
+6\gamma( R')^2\right]+
\frac{2(1+a-aR_E^2+bR_E^4-2bR_E^2)
(2\alpha+6\gamma R)
 R'}{R_E},\\[8pt]
S_2 &=
\frac{(2bR_E^2-a+ R_E-1)
(1+2\alpha R_E+3\gamma R_E^2)}{2}
+\frac{1}{R_E^2},\quad O= (S_1-S_2)\left[A(S_1-S_2)+B\right]\\[8pt]
S_3&=-(1-a R_C^2+bR_C^4)
\Big[(2\alpha+6\gamma R) R''
+6\gamma( R')^2\Big]+\frac{2(1+a-aR_C^2+bR_C^4-2bR_C^2)
(2\alpha+6\gamma R) R'}{R_C},\\[8pt]
S_4&=
\frac{(2bR_C^2-a+ R_C-1)
(1+2\alpha R_C+3\gamma R_C^2)}{2}
+\frac{1}{R_C^2}.
\end{aligned}
\]
In $S_1$ and $S_2$ the Ricci scalar and its derivatives are evaluated at $r=R_E$, and in $S_3$ and $S_4$ at $r=R_C$. The combinations $S_1-S_2$ and $S_3-S_4$ are simply the density at the surface and at the core-envelope interface, respectively. With these relations, the model depends only on $M$, $R_C$, $R_E$ and the remaining free parameters, which can then be fixed by observational data for specific compact stars.
\section{Graphical analysis}
\label{sec:4}
The solution obtained above has to pass stability tests before it can be accepted as a realistic model of a compact star. Since several of the expressions involve integrals that can only be evaluated numerically. We examine the behavior of the model through graphical analysis. For all figures, we take $R_C=5.05$~km for the core radius and $R_E=9.97$~km for the stellar radius, together with the parameter values listed in tab. \ref{tab:1} and \ref{tab:2}. The core and the envelope are shown in different colors, so that the behavior on each side of the interface can be followed separately.

\subsection{Metric potentials}\label{4.1}
A physically acceptable interior solution must have metric potentials that are finite, positive and free of singularities everywhere inside the star. At the center, $e^{-\lambda}$ should equal unity and $e^{\nu}$ should take a finite, non-zero value, and both functions should have vanishing slope there. Moving outwards, $e^{\nu}$ is expected to increase and $e^{-\lambda}$ to decrease, and at the surface both must agree with the exterior Schwarzschild metric.
\begin{figure}[!htbp]
   \centering
   \includegraphics[width=80mm]{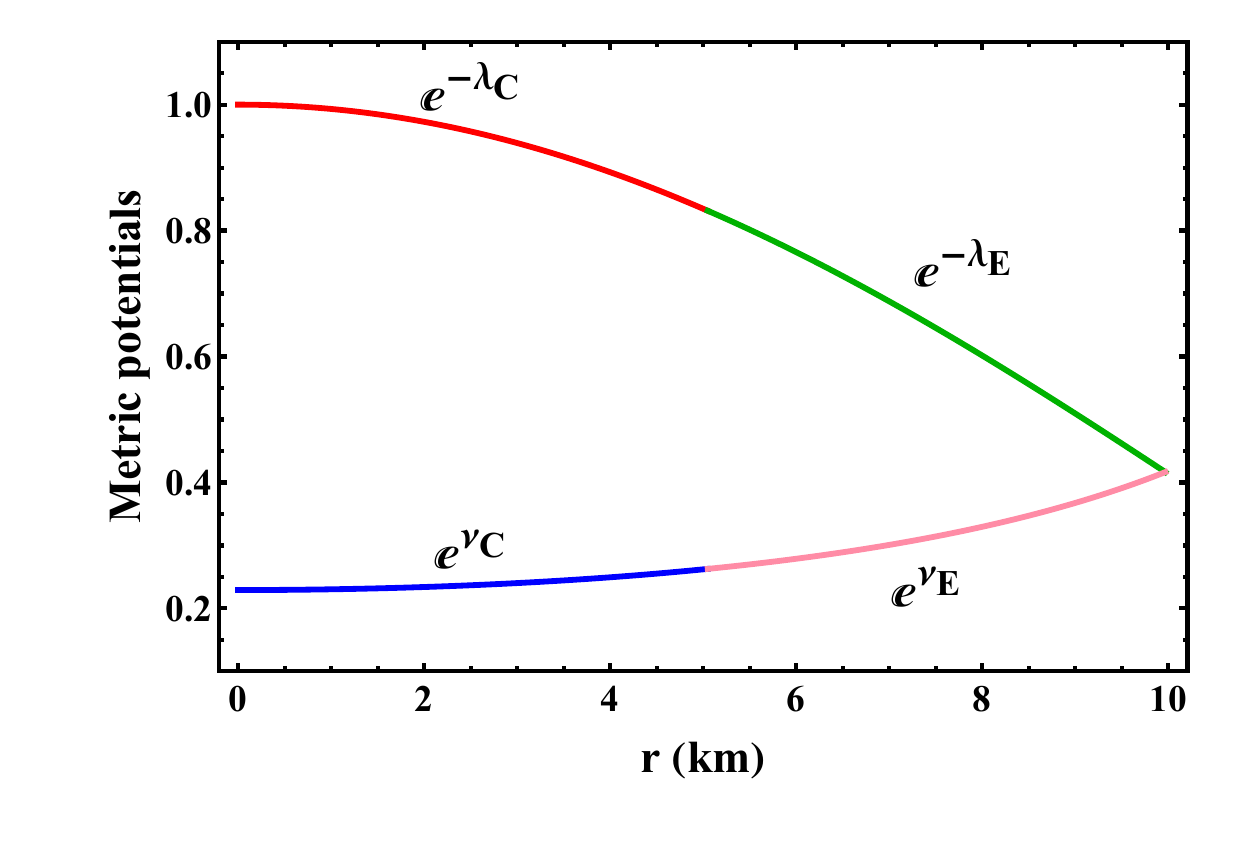}
   \caption{Radial profiles of the metric functions inside the star. The  $e^{-\lambda_C}$ (red curve) and $e^{\nu_C}$ (blue curve) in the core ($0\le r\le R_C$), and the green and pink curves show $e^{-\lambda_E}$ and $e^{\nu_E}$ in the envelope ($R_C\le r\le R_E$).}
   \label{fig:1}
\end{figure}
\FloatBarrier
Figure~\ref{fig:1} shows that our solution satisfies all of continuity requirements. The function $e^{-\lambda_C}$ starts at exactly $1$ at the center, as expected from $Z(0)=1$, and falls off slowly through the core. Both potentials are regular at $r=0$, which confirms that the geometry is uniform at the center. The two curves meet at $r=R_E$, as they must for a smooth junction with the exterior metric. At that point $e^{\nu_E}=e^{-\lambda_E}=1-(2M/R_E)$. There are no kinks at the core-envelope boundary, so the matching conditions (\ref{eq:18}) and (\ref{eq:19}) are satisfied, and the metric is well behaved for the stellar interior.

\subsection{Density and pressure profiles}\label{4.2}

\begin{figure}[!htbp]
\centering
\subfloat[]{\includegraphics[width=80mm]{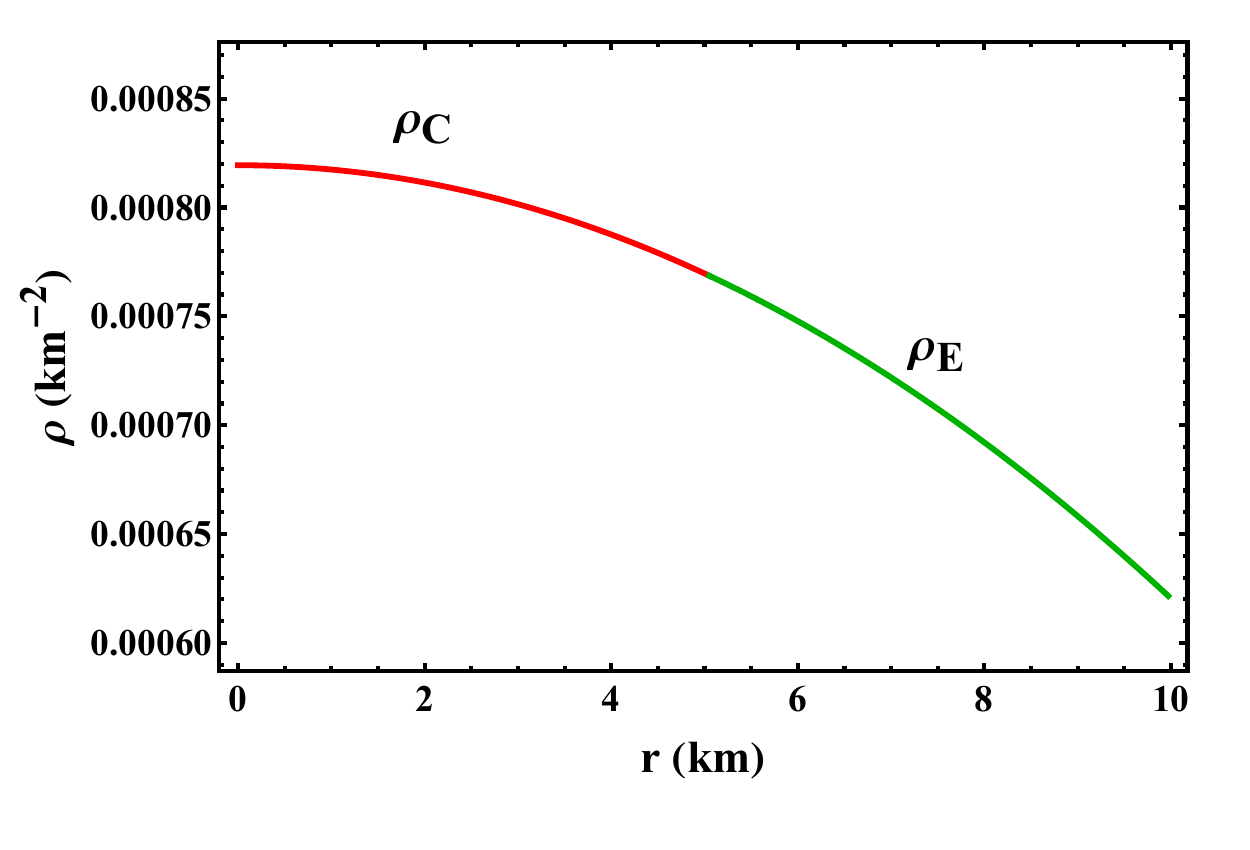}\label{fig:2a}}
\subfloat[]{\includegraphics[width=80mm]{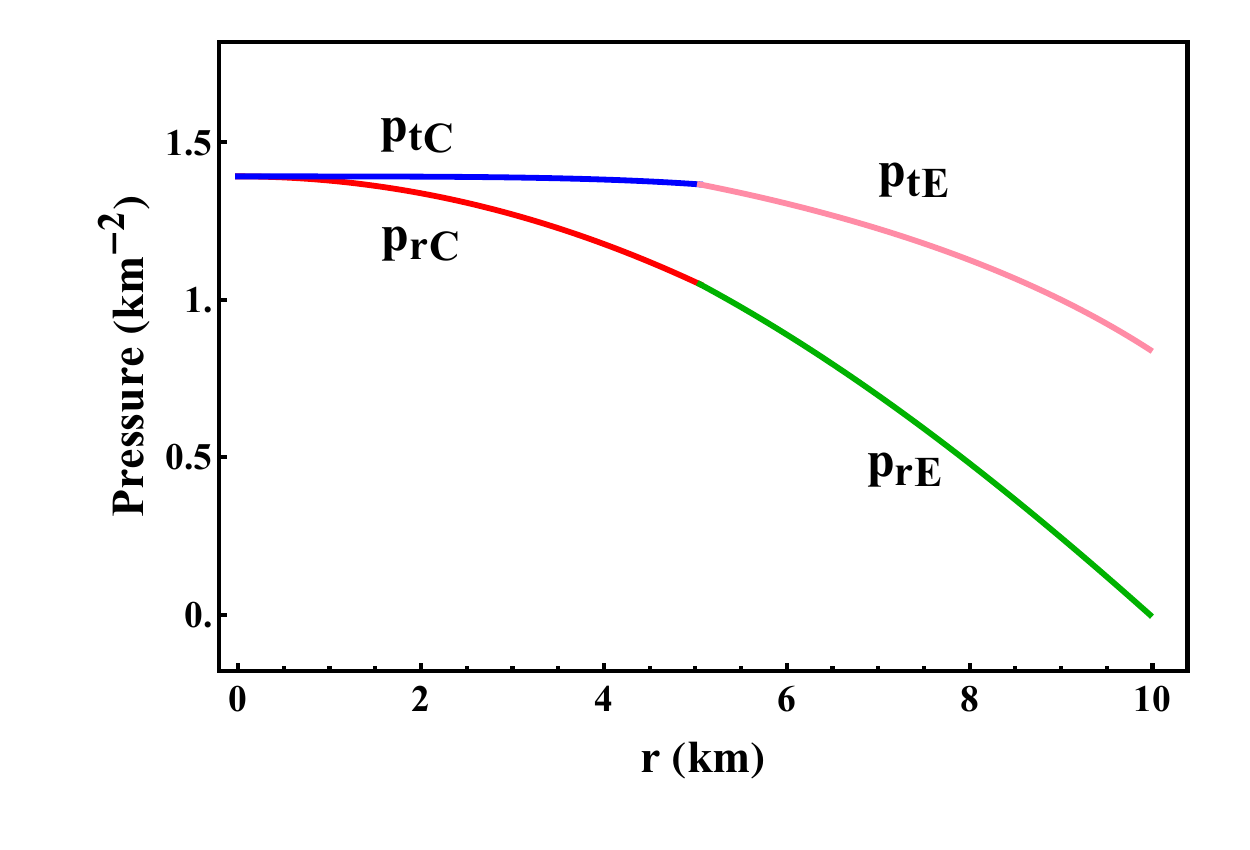}\label{fig:2b}}
\caption{(a) Density and (b) radial and tangential pressures as
functions of $r$ in the core and the envelope.}
\label{fig:2}
\end{figure}

Figure~\ref{fig:2} shows how the matter variables change with radius. The density is largest at the center, and decreases steadily towards the surface. It passes from the core to the envelope at $r=R_C$ without any jump. Both pressures also peak at the center. At far from the center the two pressures separate, with $p_t$ remaining above $p_r$ everywhere inside the star. The radial pressure falls to zero at $r=R_E$, which marks the boundary, while the $p_{t}$ stays small but non-zero. The radial pressure is continuous across the interface.
\FloatBarrier

\subsection{Anisotropy factor}\label{4.3}
\begin{figure}[!htbp]
   \centering
   \includegraphics[width=80mm]{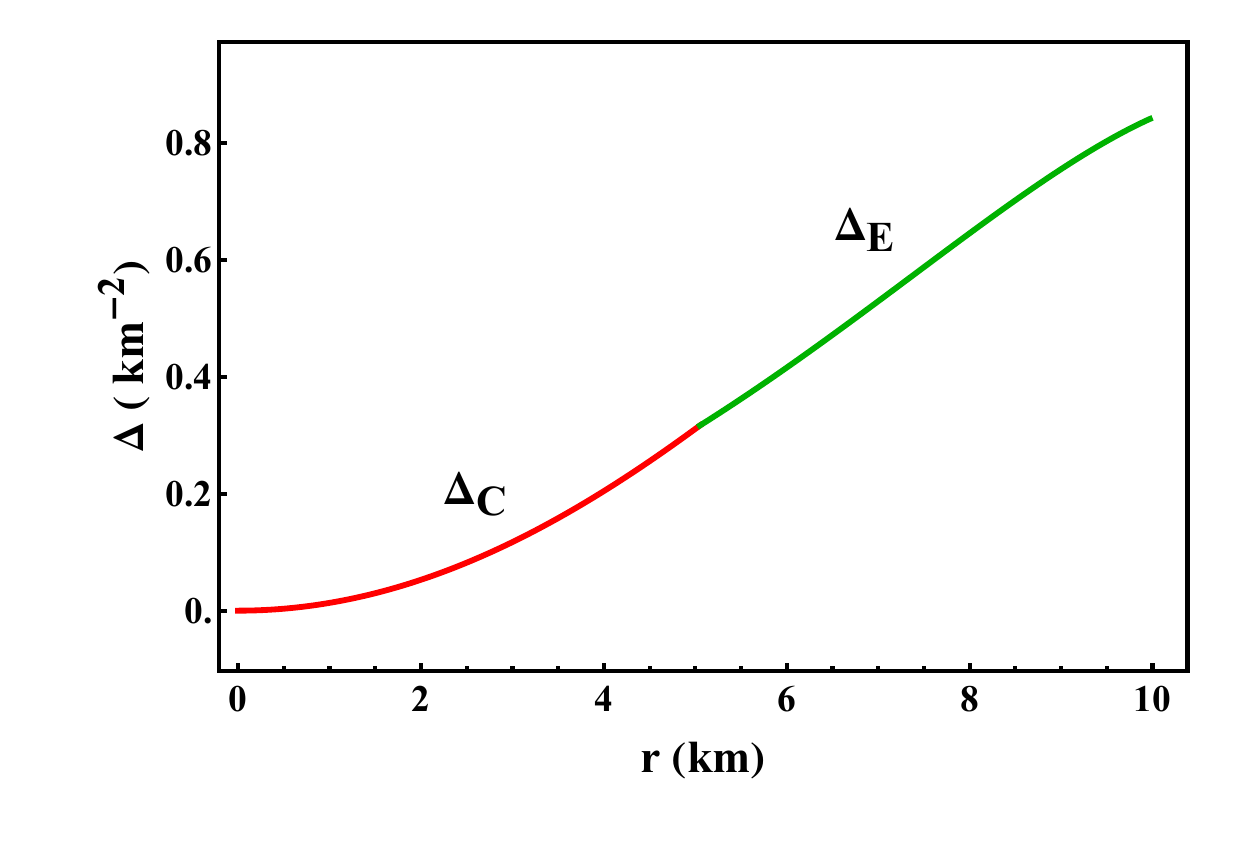}
   \caption{Anisotropy factor $\Delta$ as a function of $r$ in the core (red) and the envelope (green).}
   \label{fig:3}
\end{figure}

The anisotropy factor is plotted in Fig.~\ref{fig:3}. $\Delta$ is positive throughout the star, as the tangential pressure exceeds the radial one. The anisotropy therefore gives rise to an outward force that works against gravity and helps support the star. This extra support is one reason anisotropic models can reach higher compactness than isotropic ones.
\FloatBarrier

\subsection{Mass, compactness and gravitational redshift}\label{4.4}
\begin{figure}[!htbp]
 \centering
 \includegraphics[width=80mm]{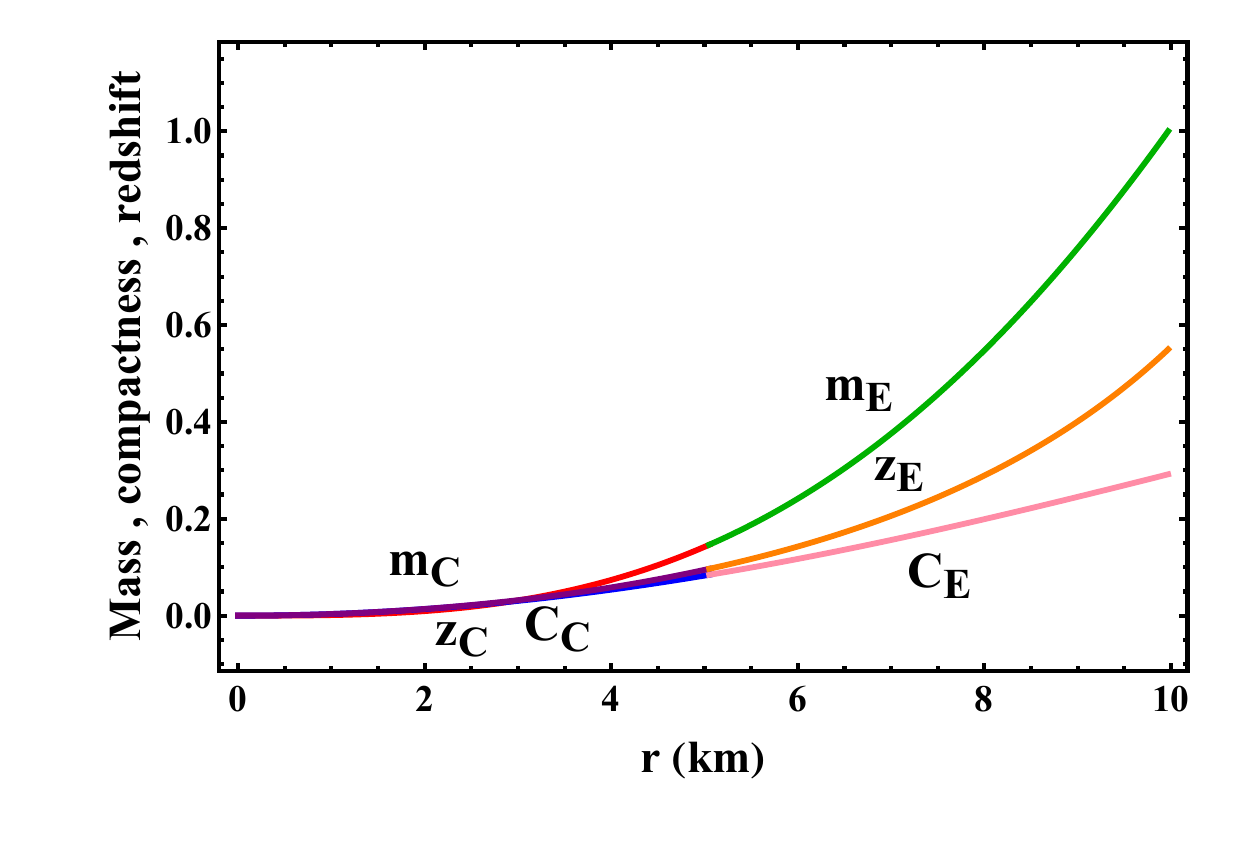}
 \caption{Mass $m$, compactness $u$ and gravitational redshift $z$ as functions of $r$ in the core and the envelope.}
 \label{fig:4}
\end{figure}

The mass enclosed within a radius $r$ is obtained by integrating the density,
\begin{equation}
m(x)=4\pi\int_{0}^{x}\rho\,x\,dr,
\label{eq:21}
\end{equation}
and the compactness and the surface redshift follow from it as
\begin{equation}
C =\frac{m}{\sqrt{x}},
\label{eq:22}
\end{equation}
and redshift is defined as 
\begin{equation}
z=\frac{1}{\sqrt{1-2C}}-1.
\label{eq:23}
\end{equation}
All three quantities are plotted in Fig.~\ref{fig:4}. Each Vanishes at the stellar center and increases steadily outwards, with no break at the core-envelope interface. The configuration is stable against gravitational collapse following the Buchdahl limit \cite{BD}.
\FloatBarrier

\subsection{Causality conditions}
\begin{figure}[!htbp]
\centering
\includegraphics[width=80mm]{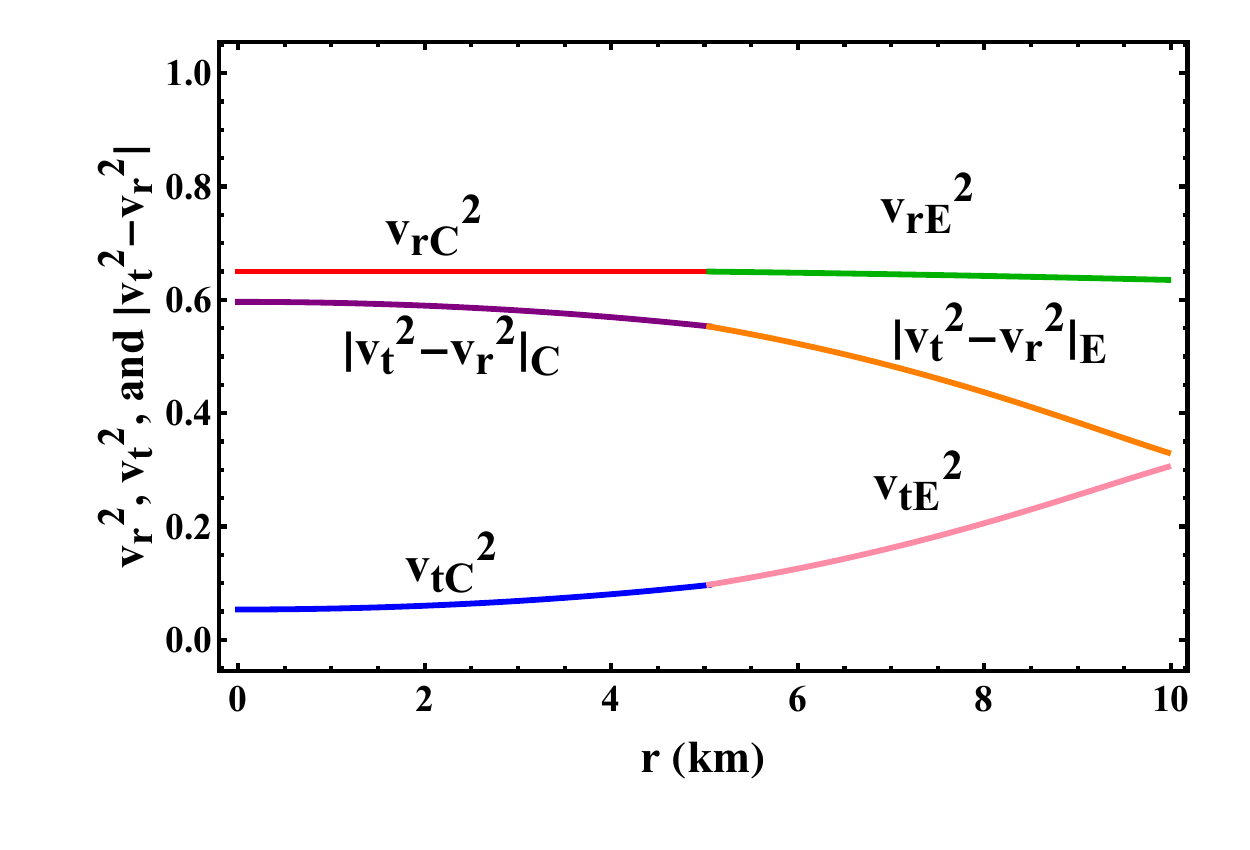}
\caption{Radial and tangential sound speeds, together with
$|v_t^2-v_r^2|$, as functions of $r$ in the core and the envelope.}
\label{fig:5}
\end{figure}

For a physically realistic model, the speed of sound inside the star must not exceed the speed of light. In units with speed of light is equal to 1, this means that both the radial and the tangential sound speeds must lie between $0$ and $1$
\begin{equation}
0\leq v_r^2=\frac{dp_r}{d\rho}\leq 1,
\qquad
0\leq v_t^2=\frac{dp_t}{d\rho}\leq 1.
\label{eq:24}
\end{equation}
Figure~\ref{fig:5} shows that both conditions are satisfied everywhere in the star. Herrera's cracking criterion \cite{H} and the condition proposed by Abreu et al. \cite{AB}, discussed the range where an anisotropic configuration is potentially stable.
\FloatBarrier
In our model $v_t^2$ remains below $v_r^2$ throughout the star, and $|v_t^2-v_r^2|$ drops from about $0.6$ at the center to about $0.33$ at the surface. The condition (\ref{eq:24}) is therefore satisfied in both the core and the envelope, and the model is stable against cracking.

\subsection{Adiabatic index}\label{4.6}

\begin{figure}[!htbp]
\centering
\includegraphics[width=80mm]{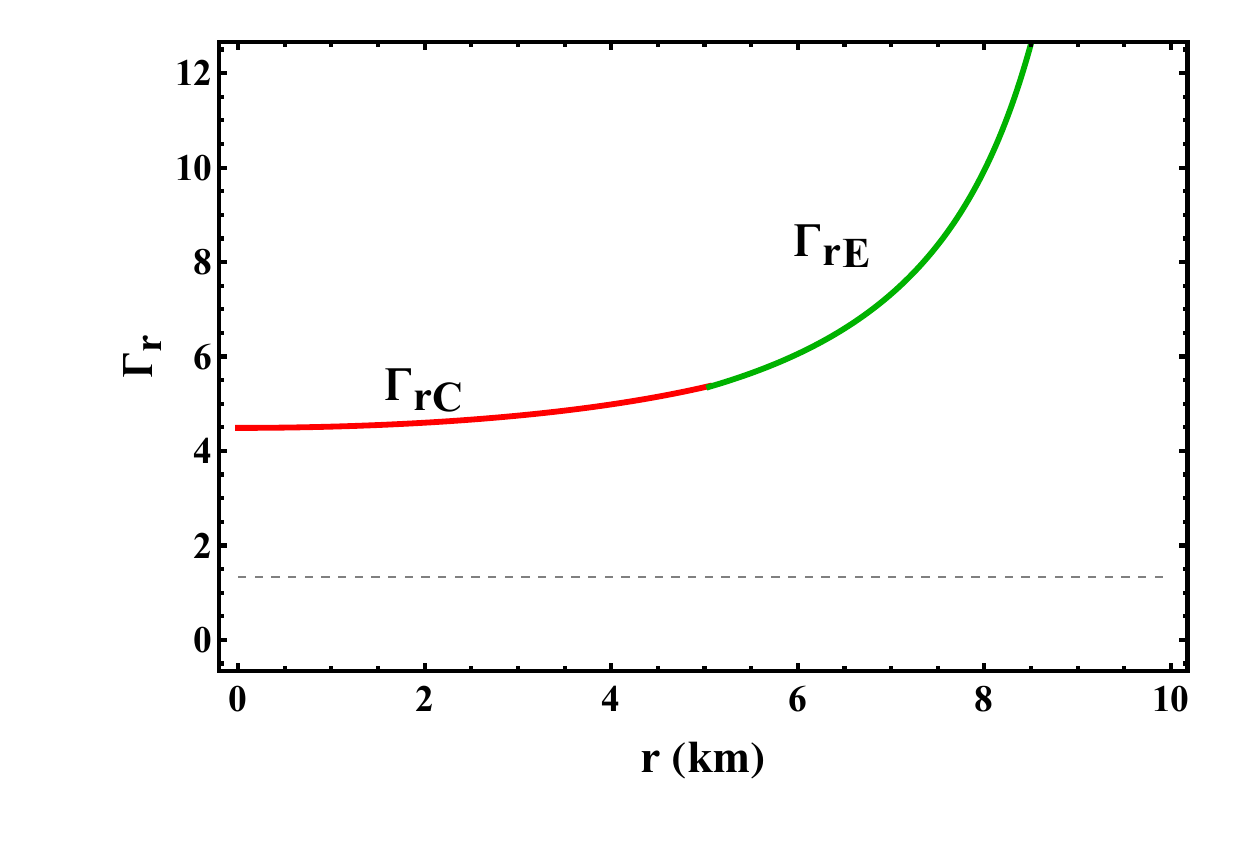}
\caption{Radial adiabatic index $\Gamma_r$ as a function of $r$ in the core (red) and the envelope (green). The dashed line marks the critical value $4/3$.}
\label{fig:6}
\end{figure}

The adiabatic index measures how stiff the EoS is, and it gives a simple test of dynamical stability. For the radial pressure it is defined as
\begin{equation}
\Gamma_r=\frac{\rho+p_r}{p_r} \frac{dp_r}{d\rho}>\frac{4}{3}.
\label{eq:25}
\end{equation}
A relativistic sphere is stable against radial perturbations only if $\Gamma_r>4/3$ \cite{BDC}. In general relativity the bound becomes somewhat stricter, but $4/3$ remains the usual reference value. As Fig.~\ref{fig:6} shows, our model stays well above this limit, marked by the dashed line, everywhere inside the star. The adiabatic index is about $4.5$ at the center and changes only a little through the core. In the envelope it rises much more steeply. This sharp growth near the surface is expected, because $\Gamma_r$ contains $1/p_r$ and the radial pressure vanishes at $r=R_E$. The model is therefore stable against radial adiabatic perturbations in both regions.
\FloatBarrier

\subsection{Energy conditions}\label{4.7}

\begin{figure}[!htbp]
\centering
\subfloat[]{\includegraphics[width=80mm]{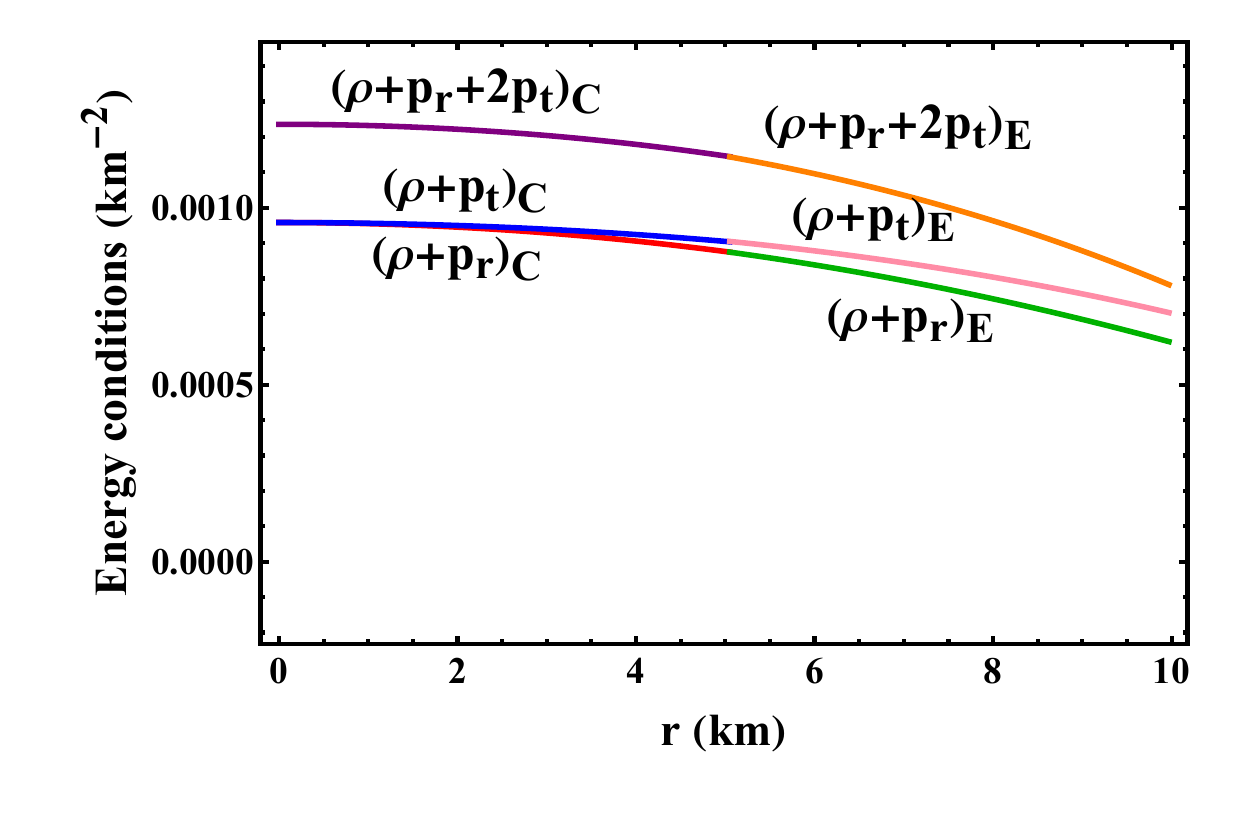}\label{fig:7a}}
\subfloat[]{\includegraphics[width=80mm]{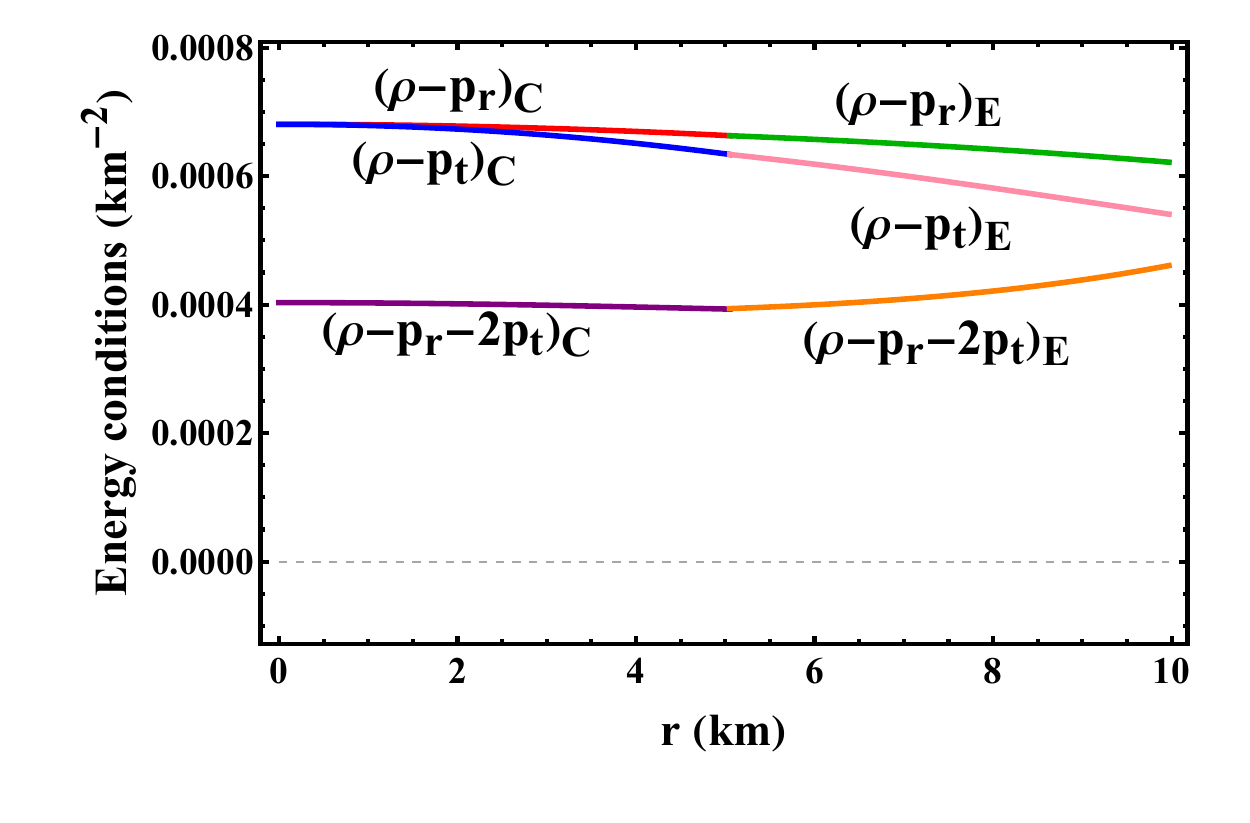}\label{fig:7b}}
\caption{Energy conditions in the core and the envelope with respect to radius.}
\label{fig:7}
\end{figure}

A physically reasonable matter distribution must also satisfy the necessary energy conditions. For an anisotropic fluid they read
\begin{itemize}
    \item $\quad \rho \geq 0,$
    \item $\quad
    (\rho+p_t \geq 0),\quad
    (\rho+p_r \geq 0),$
    \item $\quad
    \rho+2p_t+p_r \geq 0,$
    \item $\quad
    (\rho-p_t \geq 0),\quad
    (\rho-p_r \geq 0),$
    \item $\quad
    \rho-2p_t-p_r \geq 0.$
\end{itemize}
The relevant combinations are plotted in Fig.~\ref{fig:7}. Panel (a) shows that $\rho+p_t$, $\rho+p_r$ and $\rho+2p_t+p_r$ are positive throughout the star. They decrease gradually from the center to the surface but never come close to zero. Together with the positive density in Fig.~\ref{fig:2a}. Panel (b) shows the same for $\rho-p_t$, $\rho-p_r$ and $\rho-2p_t-p_r$. The first two decrease slowly outwards. All energy conditions hold in both the core and the envelope, so the matter content of the model is physically acceptable.
\FloatBarrier

\subsection{Tolman-Oppenheimer-Volkoff equation}
\begin{figure}[!htbp]
\centering
\includegraphics[width=80mm]{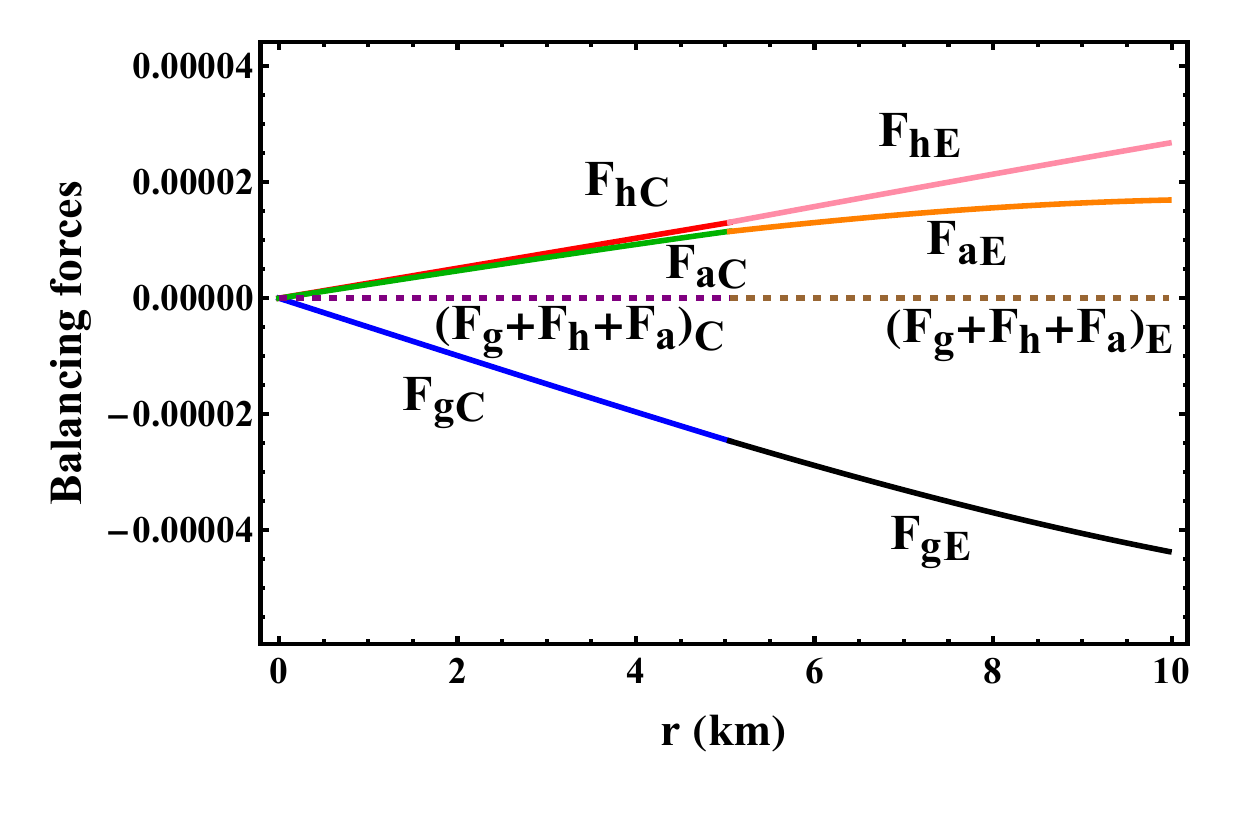}
\caption{Gravitational ($F_g$), hydrostatic ($F_h$) and anisotropic ($F_a$) forces in the core and the envelope. The dotted line shows their sum, which vanishes throughout the star.}
\label{fig:8}
\end{figure}

A stable star must also be in hydrostatic equilibrium. For an anisotropic fluid this is described by the generalized Tolman-Oppenheimer-Volkoff equation,
\begin{equation}
\underbrace{
-\frac{M_g(r)(\rho+p_r)}{r^2}
e^{\frac{\lambda-\nu}{2}}
}_{F_g}
\underbrace{
-\frac{dp_r}{dr}
}_{F_h}
+
\underbrace{
\frac{2\Delta(r)}{r}
}_{F_a}
=0,
\label{eq:26}
\end{equation}
where $M_g$ is the gravitational mass within radius $r$. The three terms in Eq.~(\ref{eq:26}) are the gravitational force $F_g$, the hydrostatic force $F_h$ and the anisotropic force $F_a$. Equilibrium requires them to cancel.
\begin{equation}
F_g+F_h+F_a=0.
\label{eq:27}
\end{equation}
In metric $f(R)$ gravity the matter energy-momentum tensor is still conserved, so the Eq. (\ref{eq:26}) keeps its familiar form, even though the metric functions themselves carry the influence of the higher-curvature terms. The three forces are plotted in Fig.~\ref{fig:8}. All of them vanish at the center and grow outwards. The gravitational force is negative, pulling matter inward. It is balanced by the hydrostatic and anisotropic forces, which are both positive and push outward. In the core these two outward forces are of similar size. In the envelope the hydrostatic force grows faster, while the anisotropic contribution starts to level off. Their sum cancels $F_g$ at every point, so the total force, shown by the dotted line, stays at zero throughout the star. The model is therefore in equilibrium in both the core and the envelope.
\FloatBarrier

\subsection{Tidal Love number and tidal deformability}

In $f(R)$ gravity the quadrupolar tidal response cannot be computed from the general relativistic Hinderer equation alone, because the additional scalar degree of freedom $f_R$ couples to the even-parity metric perturbation. We therefore follow Yazadjiev et al. \cite{TLN1} and use the dynamical equivalence of $f(R)$ gravity with a massive scalar-tensor theory. In this description the perturbation equations take their simplest form in the Einstein frame. For $f(R)=R+\alpha R^{2}+\gamma R^{3}$ we define
\begin{equation}
\Phi\equiv f_R = 1+2\alpha R+3\gamma R^{2},\qquad f_{RR}=2\alpha+6\gamma R ,
\label{eq:28}
\end{equation}
and require $\Phi>0$ and $f_{RR}>0$ throughout the configuration. These conditions exclude ghost and tachyonic instabilities. The Einstein-frame metric, scalar field and potential are
\begin{equation}
\tilde g_{\mu\nu}=\Phi\, g_{\mu\nu},\qquad \varphi=\frac{\sqrt{3}}{2}\ln\Phi,\qquad V(\varphi)=\frac{R f_R-f}{f_R^{2}}=\frac{\alpha R^{2}+2\gamma R^{3}}{\Phi^{2}} .
\label{eq:29}
\end{equation}
Derivatives with respect to $\varphi$ follow from $d\varphi/dR=\sqrt{3}\,f_{RR}/(2\Phi)$:
\begin{equation}
V_{\varphi}\equiv\frac{dV}{d\varphi} =\frac{2\left(R-\gamma R^{3}\right)}{\sqrt{3}\,\Phi^{2}}, \qquad V_{\varphi\varphi}\equiv\frac{d^{2}V}{d\varphi^{2}} =\frac{4}{3}\left[\frac{1-3\gamma R^{2}}{\Phi\, f_{RR}} -\frac{2\left(R-\gamma R^{3}\right)}{\Phi^{2}}\right].
\label{eq:30}
\end{equation}
Expanding about $R=0$ gives the mass of the scalar mode.
\begin{equation}
m_{\varphi}^{2}=\tfrac14 V_{\varphi\varphi}\big|_{R=0}=\frac{1}{6\alpha}, \qquad \lambda_{\varphi}=\frac{2\pi}{m_{\varphi}} .
\label{eq:31}
\end{equation}
The field equations follows the line element
\begin{equation}
d\tilde s^{2}=e^{2\psi}dt^{2}-e^{2L}ds^{2}-s^{2}d\Omega^{2},
\label{eq:32}
\end{equation}
we obtain
\begin{align}
s&=r\sqrt{\Phi},\qquad
e^{2\psi}=\Phi\,e^{\nu},\qquad
e^{2L}=\frac{e^{\lambda}}{\left(1+\dfrac{r\,\Phi_{,r}}{2\Phi}\right)^{2}},
\label{eq:33}\\
\rho^{*}&=\frac{\rho}{\Phi^{2}},\qquad
p^{*}=\frac{p_r}{\Phi^{2}},\qquad
\varphi_{0}=\frac{\sqrt{3}}{2}\ln\Phi ,
\label{eq:34}
\end{align}
where $e^{-\lambda}=Z=bx^{2}-ax+1$ with $x=r^{2}$, and $\Phi_{,r}=f_{RR}\,R_{,r}$ with $R(r)$ given by Eq.~(\ref{eq:13}). In what follows a prime denotes $d/ds$, related to the Jordan-frame radial derivative by
\begin{equation}
\frac{d}{ds}=\frac{1}{\mathcal{K}}\frac{d}{dr},\qquad
\mathcal{K}\equiv\frac{ds}{dr}=\sqrt{\Phi}\left(1+\frac{r\,\Phi_{,r}}{2\Phi}\right),
\label{eq:35}
\end{equation}
so that, for example,
\begin{equation}
\psi'=\frac{1}{\mathcal{K}}\left(\frac{\nu_{,r}}{2}+\frac{\Phi_{,r}}{2\Phi}\right),\qquad \varphi_{0}'=\frac{\sqrt{3}}{2}\,\frac{\Phi_{,r}}{\mathcal{K}\,\Phi}.
\label{eq:36}
\end{equation}
The Einstein-frame mass function is $\tilde m(s)=\tfrac{s}{2}\left(1-e^{-2L}\right)$, and the gravitational mass $M$ is the same in both frames~\cite{TLN1}. An external tidal field induces a static even-parity $\ell=2$ perturbation. In the Regge-Wheeler gauge it is described by the metric function $H(s)$ and the scalar-field perturbation $\delta\varphi(s)$. These satisfy the coupled system~\cite{TLN1}
\begin{align}
H''+P\,H'+Q\,H+S\,\delta\varphi&=0,
\label{eq:37}\\
\delta\varphi''+P_{\varphi}\,\delta\varphi'+Q_{\varphi}\,\delta\varphi+S_{\varphi}\,H&=0,
\label{eq:38}
\end{align}
with coefficients
\begin{align}
P&=\frac{2}{s}+e^{2L}\left[\frac{1-e^{-2L}}{s}+4\pi s\left(p^{*}-\rho^{*}\right)-\frac{s}{2}V\right],
\label{eq:39}\\
Q&=-\frac{6\,e^{2L}}{s^{2}}
+4\pi e^{2L}\left[5\rho^{*}+9p^{*}+\frac{\rho^{*}+p^{*}}{v_r^{2}}-\frac{V}{4\pi}\right]
-4\psi'^{2},
\label{eq:40}\\
S&=e^{2L}\Bigg\{-4s\,\varphi_{0}'\left[\frac{1-e^{-2L}}{s^{2}}+8\pi p^{*}
+e^{-2L}\varphi_{0}'^{2}-\frac{V}{2}\right]
-\frac{16\pi}{\sqrt{3}}\left[\rho^{*}-3p^{*}+\left(\rho^{*}+p^{*}\right)\Xi\right]
+V_{\varphi}\Bigg\},
\label{eq:41}\\
P_{\varphi}&=\psi'-L'+\frac{2}{s},
\label{eq:42}\\
Q_{\varphi}&=-e^{2L}\left\{\frac{6}{s^{2}}+4e^{-2L}\varphi_{0}'^{2}
+\frac14 V_{\varphi\varphi}
-\frac{8\pi}{3}\left[-2\left(\rho^{*}-3p^{*}\right)+\left(\rho^{*}+p^{*}\right)\Xi\right]\right\},
\label{eq:43}\\
S_{\varphi}&=e^{2L}\left\{-2e^{-2L}\psi'\varphi_{0}'
-\frac{4\pi}{\sqrt{3}}\left[\rho^{*}-3p^{*}+\left(\rho^{*}+p^{*}\right)\Xi\right]
+\frac14 V_{\varphi}\right\},
\label{eq:44}
\end{align}
where
\begin{equation}
\Xi\equiv\frac{1-3v_r^{2}}{2v_r^{2}},\qquad v_r^{2}=\frac{dp_r}{d\rho}
\label{eq:45}
\end{equation}
is the Jordan-frame (physical) radial sound speed. $V$, $V_\varphi$ and $V_{\varphi\varphi}$ are evaluated on the background through $R(r)$ via Eqs.~(\ref{eq:29})-(\ref{eq:30}), so no inversion $R(\varphi)$ is needed. As in the general relativistic treatment, the anisotropic stress $\Delta=p_t-p_r$ is neglected in the perturbation equations and $p^{*}$ is built from $p_r$. 
\paragraph{Core region ($0\le r\le R_C$).}
For $p_r=\xi\rho+\beta$:
\begin{equation}
v_{rC}^{2}=\xi,\qquad
\rho^{*}+p^{*}=\frac{(1+\xi)\rho_C+\beta}{\Phi^{2}},\qquad
\frac{\rho^{*}+p^{*}}{v_{rC}^{2}}=\frac{(1+\xi)\rho_C+\beta}{\xi\,\Phi^{2}},\qquad
\Xi_C=\frac{1-3\xi}{2\xi}.
\label{eq:46}
\end{equation}
\paragraph{Envelope region ($R_C\le r\le R_E$).}
For $p_r=A\rho^{2}+B\rho-O$:
\begin{equation}
v_{rE}^{2}=2A\rho_E+B,\qquad
\rho^{*}+p^{*}=\frac{A\rho_E^{2}+(B+1)\rho_E-O}{\Phi^{2}},\qquad
\Xi_E=\frac{1-3\left(2A\rho_E+B\right)}{2\left(2A\rho_E+B\right)}.
\label{eq:47}
\end{equation}
\paragraph{Core-envelope interface.}
Both regions use the same metric function $Z$, so the density is continuous at $r=R_C$ ($\rho_C(R_C)=\rho_E(R_C)$). Only $v_r^{2}$ jumps there, and it jumps by a finite amount. The coefficients (\ref{eq:39})-(\ref{eq:44}) therefore contain no $\delta$-function terms at the interface, and $H$, $H'$, $\delta\varphi$ and $\delta\varphi'$ are continuous across $r=R_C$. The integration simply switches from Eq.~(\ref{eq:46}) to Eq.~(\ref{eq:47}). Regularity at the center requires
\begin{equation}
H=h_{0}\,s^{2}\left[1+\mathcal{O}(s^{2})\right],\qquad
\delta\varphi=q_{0}\,s^{2}\left[1+\mathcal{O}(s^{2})\right].
\label{eq:48}
\end{equation}
The system of Eq.~(\ref{eq:37})-(\ref{eq:38}) is linear and has two independent regular solutions. We integrate it from the center to the surface twice, with $(h_0,q_0)=(1,0)$ and $(h_0,q_0)=(0,1)$, and form the general regular solution as a linear combination of the two. The overall normalization is irrelevant. The ratio $q_0/h_0$ is fixed by requiring $\delta\varphi$ to decay outside the star; the scalar field is massive, so its tidal Love number vanishes~\cite{TLN1}. Note that Eq.~(\ref{eq:37}) cannot be reduced to a closed first-order Riccati equation for $y=sH'/H$, because the source term $S\,\delta\varphi/H$ is not a known function of $s$. The two second-order equations must be integrated together.

\paragraph{Exterior and matching.}
The junction conditions (\ref{eq:20}) with the Schwarzschild exterior (\ref{eq:17c}) require $[R]^{+}_{-}=0$, and in the absence of a thin shell also $[\partial_{r}R]^{+}_{-}=0$~\cite{TLN7}. Hence $R(R_E)=0$, $\Phi(R_E)=1$, $\varphi_{0}(R_E)=0$ and $s(R_E)=R_E$. These conditions must be verified explicitly for the chosen parameters. In the exterior, $\rho^{*}=p^{*}=\varphi_0=V=V_{\varphi}=0$. Then $S=S_\varphi=0$: $H$ obeys the general-relativistic equation~\cite{TLN6}, and $\delta\varphi$ decouples:
\begin{equation}
\delta\varphi''+\frac{2(r-M)}{r(r-2M)}\,\delta\varphi'
-\left(1-\frac{2M}{r}\right)^{-1}\left(\frac{6}{r^{2}}+m_{\varphi}^{2}\right)\delta\varphi=0 .
\label{eq:49}
\end{equation}
Its solution, $\delta\varphi\sim e^{-m_{\varphi}r}/r$ as $r\to\infty$, is obtained by integrating Eq.~(\ref{eq:49}) inward. It defines $\kappa\equiv\delta\varphi'/\delta\varphi|_{r=R_E^{+}}$.

\paragraph{Surface density discontinuity.}
The envelope EoS vanishes at a non-zero density $\rho_s$, given by $A\rho_s^{2}+B\rho_s-O=0$, so $\rho$ drops from $\rho_s$ to $0$ at $r=R_E$. The terms proportional to $1/v_r^{2}=d\rho/dp_r$ in $Q$, $S$, $Q_{\varphi}$ and $S_{\varphi}$ then contain $\delta$-functions. Using the background equilibrium equation $dp_r/ds=-(\rho+p_r)\left(\psi'-\varphi_0'/\sqrt{3}\right)$, they produce the jumps
\begin{equation}
\big[H'\big]^{+}_{-}=\mathcal{J}\left(H-\frac{2}{\sqrt{3}}\,\delta\varphi\right),\qquad \big[\delta\varphi'\big]^{+}_{-}=\mathcal{J}\left(\frac{\delta\varphi}{3}-\frac{H}{2\sqrt{3}}\right),
\label{eq:50}
\end{equation}
\begin{equation}
\mathcal{J}=-\frac{4\pi e^{2L}\rho_s}{\Phi^{2}\left(\psi'-\varphi_{0}'/\sqrt{3}\right)}\Bigg|_{R_E}
=-\frac{4\pi R_E^{2}\rho_s}{M},
\label{eq:51}
\end{equation}
where the second equality uses $\Phi(R_E)=1$, $\varphi_0'(R_E)=0$ and $p_r(R_E)=0$. In the general-relativistic limit ($\delta\varphi\to0$), Eq.~(\ref{eq:50}) reduces to the familiar correction $y\to y-4\pi R_E^{3}\rho_s/M$~\cite{TLN7}. The ratio $q_0/h_0$ is fixed by the linear condition $\delta\varphi'(R_E^{+})=\kappa\,\delta\varphi(R_E^{+})$, applied after the jump (\ref{eq:50}). The tidal parameter is then
\begin{equation}
y_R=\frac{R_E\,H'(R_E^{+})}{H(R_E)} .
\label{eq:52}
\end{equation}

\subsubsection{Love number and tidal deformability}

Matching to the exterior solution $H=c_1 r^{-3}\left[1+\mathcal{O}(M/r)\right]+c_2 r^{2}\left[1+\mathcal{O}(M/r)\right]$ gives~\cite{TLN6,TLN1}
\begin{align}
&k_2=\frac{8C_1^{5}}{5}\,
\frac{\left[2C(-1+y)+2-y\right] (1-2C)^{2}}{\mathcal{D}(C,y)},
\label{eq:53}\\
&\mathcal{D}(C,y)=4C^{3}\left[13+C(-2+3y)-11y+2C^{2}(y+1)\right]
+2C\left[6+3C(5y-8)-3y\right]
\nonumber\\
&\quad+3(-2C+1)^{2}\left[2+2C(-1+y)-y\right]\ln(-2C+1),
\label{eq:54}
\end{align}
and
\begin{equation}
\Lambda=\frac{2}{3}\,k_2\,C_1^{-5}
=\frac{16}{15}\,\frac{(1-2C)^{2}\left[2+2C(y-1)-y\right]}{\mathcal{D}(C,y)} .
\label{eq:55}
\end{equation}
For the Schwarzschild-matched configuration considered here the matching radius is the stellar surface, so $y=y_R$ and $C=M/R_E$. If $R(R_E)\neq0$, the exterior is not Schwarzschild and the scalar field extends beyond the surface. One must then construct the exterior background, integrate (\ref{eq:38})-(\ref{eq:39}) out to a radius $r_{\rm match}$ several Compton wavelengths $\lambda_{\varphi}$ beyond the surface, where $\varphi_0$ and $\delta\varphi$ are negligible, choose $q_0/h_0$ so that $\delta\varphi(r_{\rm match})\to0$, and use $y=r_{\rm match}H'/H$, $C=M/r_{\rm match}$ and $C_1=M/R_S$ in Eqs.~(\ref{eq:53})-(\ref{eq:55})~\cite{TLN1}.
\begin{figure}[!htbp]
\centering
\subfloat[]{\includegraphics[width=70mm]{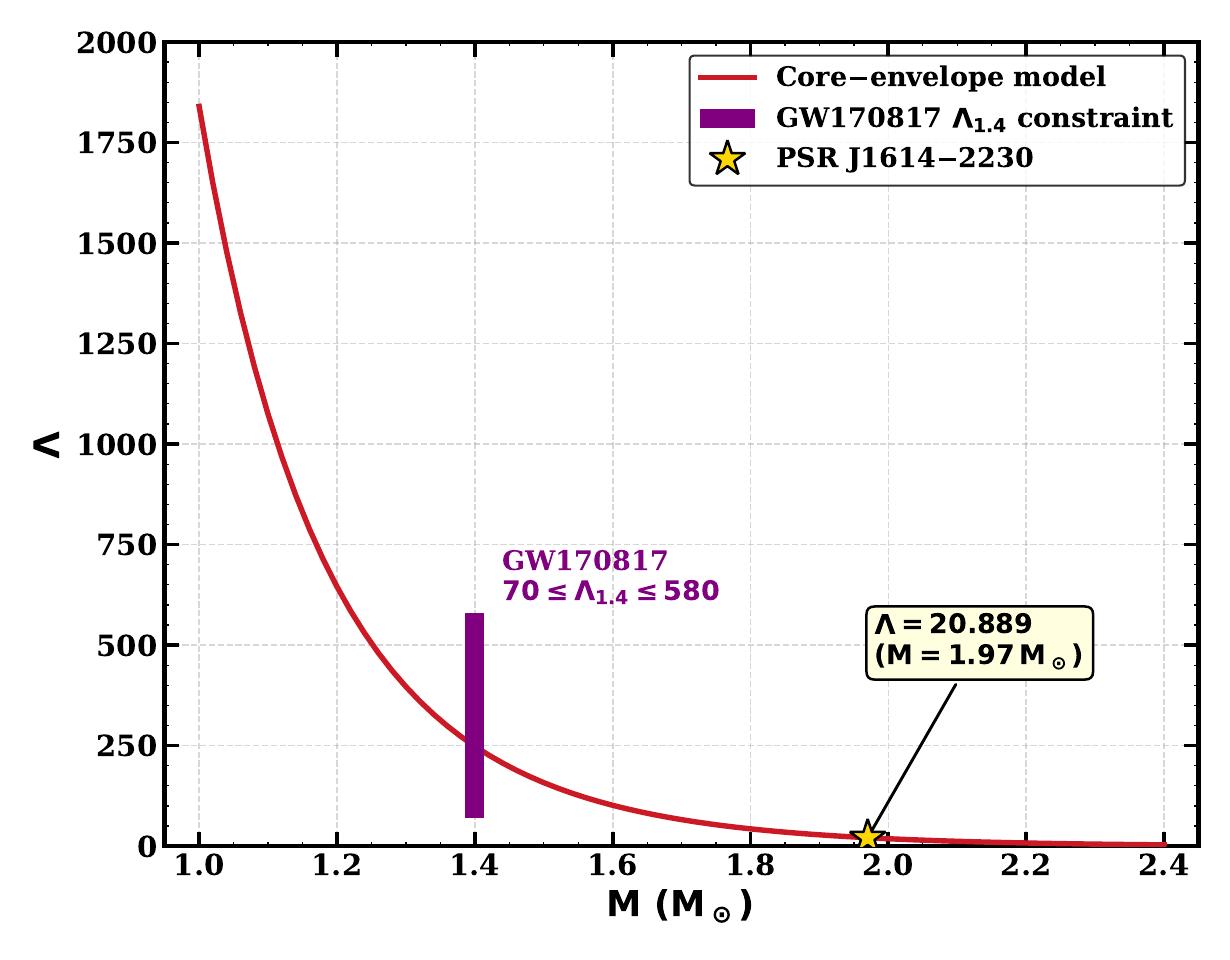}\label{fig:9a}}
\subfloat[]{\includegraphics[width=70mm]{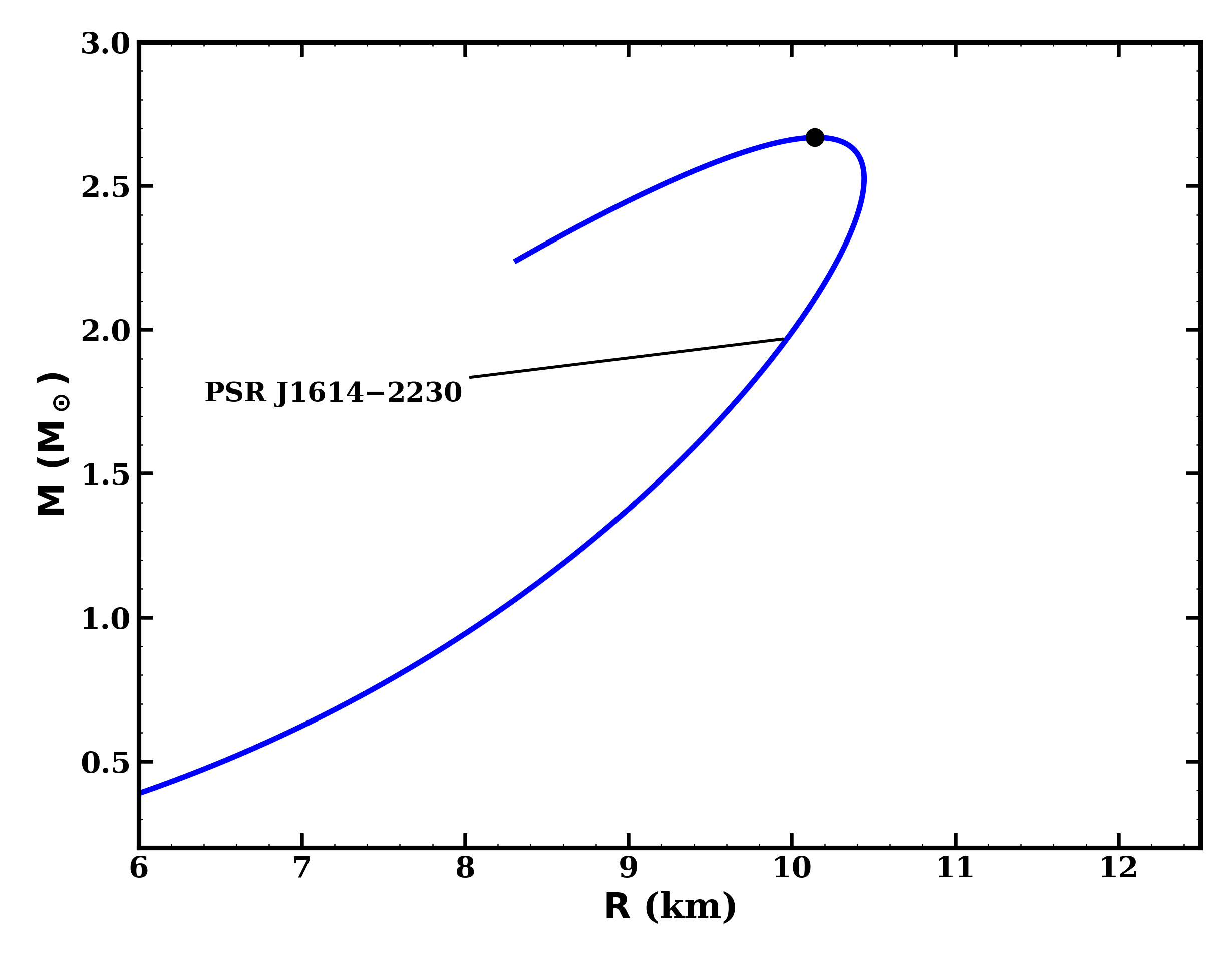}\label{fig:9b}}
\hspace{200mm}
\caption{Plots are (a) $\Lambda$ versus $M$ with the GW170817 bound; (b) mass–radius relation for PSR J1614$-$2230.}
\label{fig:9}
\end{figure}
\FloatBarrier
Figure~\ref{fig:9a} shows that the tidal deformability $\Lambda$ falls steeply as the mass increases. At $1.4\,M_\odot$ it is about $250$, which lies inside the GW170817 bound $70\le\Lambda_{1.4}\le580$ \cite{111}, and at $1.97\,M_\odot$ it drops to $\Lambda\approx20.9$. The mass-radius curve in Fig.~\ref{fig:9b} shows maximum mass for neutron star PSR J1614$-$2230.

\section{Conclusions}
\label{sec:5}

In this study, we have developed a two layered compact star model for a massive pulsar PSR J1614$-$2230, within the modified gravity $f(R)=R+\alpha R^2+\gamma R^3$ framework. The model consisting of two layers, a dense core and an outer envelope, each described by its own EoS and corresponding equations. The core is described by the linear EoS, and the envelope by the quadratic equation with the Tolman VII approximation is used in both regions for metric potentials in both regions. The model's continuity at the core-envelope interface is ensured by matching the envelope to the Schwarzschild exterior through the junction conditions. We then examined how this stellar configuration deforms under the tidal field. The tidal perturbation is then governed by two coupled equations, for the metric perturbation $H$ and a scalar perturbation equation $\delta\varphi$. For pulsar PSR J1614$-$2230, we have $M=1.97\,M_\odot$, $R_C=5.05$~km, $R_E=9.97$~km, $k_2\approx0.066$ and $\Lambda\approx20.9$. This star gives $\Lambda_{1.4}\approx250$, which lies within the range allowed by GW170817. When the modified gravity constant $\alpha$ is increased from $0$ to $1$~km$^2$, $\Lambda$ changes from $20.906$ to $20.886$, which shows that the tidal response is affected by the matter distribution rather than only by the curvature terms.\\

The graphical analysis show that the metric potentials are regular and join the exterior solution smoothly. The density and pressures decrease monotonically from the center to boundary, while the anisotropy is positively increasing . The energy conditions, causality, the cracking criterion and the adiabatic stability bound hold in both layers. The hydrostatic equilibrium is observed to hold by the balance of gravitational, hydrostatic, and anisotropic forces throughout the interior of star. The mass-radius curve passes through the observed mass of PSR J1614$-$2230 on its stable branch. The surface compactness, $C\approx0.29$, and the gravitational redshift, $z\approx0.55$, also remain within the accepted limits for anisotropic compact stars. Table~\ref{tab:1} lists the parametric values and tidal variable along the stellar radius. Table~\ref{tab:2} shows how the model and its tidal properties change with the coupling $\alpha$. Overall, the values listed in Tables~\ref{tab:1} and ~\ref{tab:2} show that the linear-quadratic core-envelope model in cubic $f(R)$ gravity remains regular, stable, properly matched to the exterior spacetime and it gives a tidal deformability for PSR J1614$-$2230 consistent with current observational limits. 

\begin{table*}[t]
\centering
\caption{Variation of $\rho$, $p_r$, $e^{\lambda}$, $\mathcal{P}(r)$, $\mathcal{Q}(r)$,
$dy/dr$ and $y(r)$ along the integration path for PSR J1614--2230
($M=1.97\,M_\odot$, $R_E=9.97$ km, $R_C=5.05$ km \cite{PSR5}) with
$b=10^{-5}$ km$^{-4}$, $a=2M/R_E^{3}+bR_E^{2}=0.006865$ km$^{-2}$,
$\xi=0.65$, $A=50$ km$^{2}$ and $B=0.65$. Densities and pressures in km$^{-2}$.}
\label{tab:1}
\setlength{\tabcolsep}{2pt}
\begin{tabular}{llrrrrrrr}
\toprule
$r$ (km) & Region & $\rho\times10^{4}$ & $p_r\times10^{4}$ & $e^{\lambda}$ & $\mathcal{P}$ & $\mathcal{Q}$ & $dy/dr$ & $y(r)$ \\
\midrule
0.05 & Core & 8.19395 & 1.387554 & 1.00002 & 39.9999 & -2399.95 & -0.000588 & 1.999985 \\
0.50 & Core & 8.18903 & 1.384353 & 1.00172 & 3.99915 & -23.9555 & -0.005873 & 1.998531 \\
1.00 & Core & 8.17410 & 1.374655 & 1.00690 & 1.99830 & -5.95556 & -0.011727 & 1.994130 \\
2.00 & Core & 8.11442 & 1.335861 & 1.02806 & 0.99652 & -1.45580 & -0.023285 & 1.976601 \\
4.95 & Core & 7.70654 & 1.070736 & 1.19360 & 0.39388 & -0.20280 & -0.053174 & 1.861343 \\
5.05 & Core$\to$Env & 7.68664 & 1.057804 & 1.20273 & 0.38559 & -0.19333 & -0.053940 & 1.855987 \\
5.15 & Env & 7.66635 & 1.043057 & 1.21217 & 0.37759 & -0.18659 & -0.044405 & 1.851612 \\
7.00 & Env & 7.21917 & 0.719110 & 1.45424 & 0.26746 & -0.08713 & -0.050990 & 1.758544 \\
9.00 & Env & 6.58256 & 0.261375 & 1.96241 & 0.18886 & -0.04887 & -0.007425 & 1.686748 \\
9.90 & Env & 6.24415 & 0.019710 & 2.36261 & 0.15671 & -0.04458 & 0.053037 & 1.704176 \\
9.97 & Env & 6.21648 & 0.000000 & 2.40120 & 0.15413 & -0.04456 & 0.059713 & 1.708121 \\
\bottomrule
\end{tabular}
\end{table*}

\begin{table*}[t]
\centering
\caption{Effect of the $f(R)$ coupling $\alpha$ on the model of
PSR J1614-2230. EoS: core $p_r=\xi\rho+\beta$ with
; envelope $p_r=A\rho^2+B\rho-O$. Radial pressures in km$^{-2}$ values are listed. $y$ is evaluated at
$r_{\rm match}$, and $k_2$, $\Lambda$ follow from
Eqs.~(\ref{eq:53})-(\ref{eq:55}) with $C=M/r_{\rm match}$ and $C_1=M/R_E$.}
\label{tab:2}
\small
\setlength{\tabcolsep}{2pt}
\begin{tabular}{lllllllllllll}
\toprule
$\alpha$ (km$^2$) & $\gamma$ (km$^4$) & $p_r(0)\times10^{4}$ & $p_r(R_C)\times10^{4}$ & $v_{r,C}^{2}$ & $v^2_{r,E}$ & $r_{\rm match}$ (km) & $y$ & $k_2$ & $\Lambda$ \\
\midrule
0.0 & 0.0 & 1.286115 & 1.057804 & 0.45 & 0.7269 & 9.970 & 1.70644 & 0.066308 & 20.906 \\
0.05 & 0.02 & 1.311483 & 1.056804 & 0.50 & 0.7269 & 12.25 & 1.70700 & 0.066291 & 20.900 \\
0.10 & 0.02 & 1.336850 & 1.055804 & 0.55 & 0.7269 & 15.22 & 1.70744 & 0.066276 & 20.896 \\
0.20 & 0.02 & 1.387586 & 1.054804 & 0.60 & 0.7269 & 19.23 & 1.70781 & 0.066264 & 20.892 \\
0.50 & 0.02 & 1.412954 & 1.053804 & 0.65 & 0.7269 & 27.53 & 1.70812 & 0.066255 & 20.889 \\
1.00 & 0.02 & 1.438322 & 1.052804 & 0.70 & 0.7269 & 39.01 & 1.70838 & 0.066246 & 20.886 \\
\bottomrule
\end{tabular}
\end{table*}
\FloatBarrier




\end{document}